\documentclass[twocolumn,showpacs,english,preprintnumbers]{revtex4}
\usepackage{graphicx}
\usepackage{dcolumn}
\usepackage{bm}
\usepackage{amssymb}
\usepackage{amsmath}
\usepackage{epsfig}
\usepackage{hyperref}
\usepackage{color}

\begin{document}

\title{Stable Solitons in Three Dimensional Free Space without the Ground
State: Self-Trapped Bose-Einstein Condensates with Spin-Orbit Coupling}
\author{Yong-Chang Zhang$^{1}$}
\author{Zheng-Wei Zhou$^{1}$}
\email{zwzhou@ustc.edu.cn}
\author{Boris A. Malomed$^{2}$}
\author{Han Pu$^{3,4}$}
\email{hpu@rice.edu}
\affiliation{$^1${Key Laboratory of Quantum Information, and Synergetic Innovation Center
of Quantum Information and Quantum Physics, University of Science and
Technology of China, Hefei, Anhui 230026, China}\\
$^2${Department of Physical Electronics, School of Electrical Engineering,
Faculty of Engineering, Tel Aviv University, 69978 Tel Aviv, Israel}\\
$^3${Department of Physics and Astronomy, and Rice Center for Quantum
Materials, Rice University, Houston, TX 77005, USA}\\
$^4${Center for Cold Atom Physics, Chinese Academy of Sciences, Wuhan
430071, China}}

\begin{abstract}
By means of variational methods and systematic numerical analysis, we
demonstrate the existence of metastable solitons in three-dimensional (3D)
free space, in the context of binary atomic condensates combining contact
self-attraction and spin-orbit coupling, which can be engineered by
available experimental techniques. Depending on the relative strength of the
intra- and inter-component attraction, the stable solitons feature a
semi-vortex or mixed-mode structure. In spite of the fact that the local
cubic self-attraction gives rise to the supercritical collapse in 3D, hence
the setting produces no true ground state, the solitons are stable against small
perturbations, motion, and collisions.
\end{abstract}

\pacs{03.75.Mn, 03.75.Lm, 05.45.Yv, 42.65.Tg}
\maketitle

\emph{Introduction and model --- } Solitons result from the balance between
dispersion and nonlinearity in diverse physical systems. Stable solitons in
one dimension (1D) have been studied extensively in diverse media, most
notably nonlinear optics and atomic Bose-Einstein condensates (BECs) \cite%
{books}. Multidimensional solitons were also predicted to exist in
ferromagnets \cite{Cooper}, superconductors \cite{Babaev}, semiconductors
\cite{semi}, BECs \cite{Ruostekoski}, baryonic matter \cite{Alkofer}, field
theory \cite{field}, \textit{etc}. However, creation of 2D and 3D bright
solitons is a much more challenging problem than in 1D. The fundamental
difficulty is the fact that the ubiquitous cubic local self-attractive
nonlinearity gives rise to the critical and supercritical collapse (blowup)
in the 2D and 3D geometry, respectively \cite{Collapse1,Collapse2,Collapse3}%
, which makes all the bright solitons unstable (the self-repulsive
nonlinearity supports stable 2D dark solitons in the form of delocalized
vortices \cite{vortex}). Several theoretical schemes have been elaborated
for the stabilization of 2D and 3D solitons. They rely on the use of
trapping potentials \cite{Dodd,Morise,BBB,Baizakov,Mihalache05},
sophisticated nonlinear interactions \cite{Mihalache02,RMP,Brazil,Driben},
or nonlocal nonlinearity \cite{Pohl,Tikho}. However, it is commonly believed
that a local cubic self-attraction may never give rise to stable solitons in
3D free space \cite{RMP,dark}.

Recently, an essential result \cite{Sakaguchi}, which helps to resolve a
related but easier problem of the stabilization of solitons in 2D free space
with local cubic attraction, has been reported in the framework of the model
of a binary BEC subject to the action of spin-orbit coupling (SOC) \cite{Lin}
(solitons in 1D SOC models have been predicted too \cite{1d}, but their
stability is obvious). It was found that the system gives rise to \emph{%
completely stable} 2D bright solitons as the ground state (GS).
The stabilization is explained by the fact that the linear SOC terms come
with a coefficient whose dimension is inverse length. The usual 2D systems
without SOC feature a specific scaling invariance, which is closely related
to the critical collapse. The scaling invariance makes the family of 2D
solitons degenerate (they are called \textit{Townes solitons} in that case
\cite{Townes}), with a single value of the norm that does not depend on the
soliton's chemical potential. This norm determines the threshold for the
onset of the critical collapse \cite{Collapse1,Collapse2}. Breaking the
scaling invariance by introducing a fixed length scale leads to the
stabilization of 2D solitons. This can be achieved by adding trapping
potentials \cite{Dodd,Morise,BBB,Baizakov,Mihalache05} or, in the free
space, with the help of SOC \cite{Sakaguchi}, which creates the missing GS
by pushing the norm of the 2D solitons below the collapse threshold. A
similar mechanism enables the stabilization of 2D spatiotemporal solitons in
a planar optical coupler \cite{Barcelona}, with the coupling's temporal
dispersion \cite{Chiang} emulating the SOC effect.

It has been previously shown that, besides the stabilization of 2D solitons,
the interplay of SOC and intrinsic BEC nonlinearity give rise to a variety
of other remarkable phenomena \cite{Effect0}. However, the possibility of
stabilizing 3D solitons in free space with the help of SOC remained an open
question. The fundamental difficulty is that, on the contrary to the 2D
situation, the supercritical collapse in 3D has zero threshold, hence the
norm cannot take values below the threshold, making the stabilization
mechanism outlined above irrelevant in 3D. The present work reveals that,
nevertheless, the self-attractive binary SOC condensate can support
(meta)stable 3D solitons in free space, in spite of the fact that the
setting has \emph{no GS} at any value of the norm (in other words, the
energy is unlimited from below). We find that the SOC-induced modification
of the dispersion of the 3D condensate may balance the attractive
nonlinearity, creating metastable solitons. In addition to the absence of
the GS, another fundamental difference of this mechanism from what is
outlined above for 2D is that the stability of the 3D solitons is controlled
not by the norm, but rather by their energy.

We follow the usual mean-field approach, defining $\Psi (\mathbf{r})=(\psi
_{+},\,\psi _{-})^{T}$ as the condensate wave function, with $\pm $
referring to two pseudo-spin components. Fixing by means of rescaling the
atomic mass and Planck's constant to be $1$, we write the system's energy as
the sum of kinetic, SOC, and interaction terms:
\begin{eqnarray}
E_{\mathrm{tot}} &=&E_{\mathrm{kin}}+E_{\mathrm{soc}}+E_{\mathrm{int}}\,,
\label{eq1} \\
E_{\mathrm{kin}} &=&\frac{1}{2}\int d^{3}r\,\Psi ^{\dag }\mathbf{p}^{2}\Psi
,~E_{\mathrm{soc}}=\lambda \int d^{3}r\,\Psi ^{\dag }\left( \mathbf{p}\cdot {%
\boldsymbol{\sigma }}\right) \Psi ,  \notag \\
E_{\mathrm{int}} &=&-\frac{g}{2}\int d^{3}r\,\left( |\psi _{+}|^{4}+|\psi
_{-}|^{4}+2\eta |\psi _{+}\psi _{-}|^{2}\right) \,,  \notag
\end{eqnarray}%
where ${\boldsymbol{\sigma }}=(\sigma _{x},\sigma _{y},\sigma _{z})$ are
Pauli matrices, and $\mathbf{p}=-i\nabla $ is the momentum operator. We
adopt the 3D isotropic form of the SOC with strength $\lambda $ \cite%
{Anderson}. The intra- and inter-component interaction strengths are
defined, respectively, as $-g$ and $-\eta g$, with $g>0$ corresponding to
the self-attraction, $\eta $ being the relative cross-nonlinearity strength.
Below, we fix the nonlinearity strength, by rescaling the wave functions, to
$g=1$ and vary the SOC strength $\lambda $, norm $N$, and cross-nonlinearity
strength $\eta $.

\begin{figure}[tbp]
\centering\includegraphics[width=0.7\columnwidth]{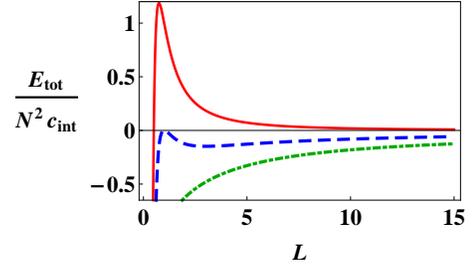}
\caption{(Color online) $E_{\mathrm{tot}}$ as a function of condensate's
size $L$, as per Eq. (\protect\ref{eq2}). The red solid, blue dashed, and
green dot-dashed lines represent the energy's variation when $\protect%
\lambda =0$, and $\protect\lambda >0$ does or does not satisfy condition (%
\protect\ref{eq3}), respectively.}
\label{fig1}
\end{figure}

\emph{Dimensional analysis ---} If $L$ is a characteristic size of the
self-trapped condensate, an estimate for the amplitudes of the wave functions
with norm $N=\int d^{3}{r}\,\left( |\psi _{+}|^{2}+|\psi
_{-}|^{2}\right) $ is $\left( \left\vert \psi _{\pm }\right\vert \right)
_{\max }\sim \sqrt{N}L^{-3/2}$. Therefore, the three terms in Eq. (\ref{eq1}%
) scale with $L$ as
\begin{equation}
E_{\mathrm{tot}}/N\sim c_{\mathrm{kin}}L^{-2}-c_{\mathrm{soc}}\lambda
L^{-1}-\left( c_{\mathrm{int}}^{\mathrm{(self)}}+c_{\mathrm{int}}^{\mathrm{%
(cross)}}\eta \right) NL^{-3}\,,  \label{eq2}
\end{equation}%
with positive coefficients $c_{\mathrm{kin}}$, $c_{\mathrm{soc}}$, and $c_{%
\mathrm{int}}^{\mathrm{(self/cross)}}$. As shown in Fig.~\ref{fig1}, Eq.~(\ref{eq2}) gives
rise to a local minimum of $E_{\mathrm{tot}}(L)$ at finite $L$, provided
that
\begin{equation}
0<\lambda N<{c_{\mathrm{kin}}^{2}}/\left[ {3\left( c_{\mathrm{int}}^{\mathrm{%
(self)}}+c_{\mathrm{int}}^{\mathrm{(cross)}}\eta \right) c_{\mathrm{soc}}}%
\right] \,.  \label{eq3}
\end{equation}%
Although this minimum cannot represent the GS (which
formally corresponds to $E_{\mathrm{tot}}\rightarrow -\infty $ at $%
L\rightarrow 0$ in the collapsed state, i.e., the system has no true GS), it
corresponds to a self-trapped state stable against small perturbations.
Previously, a similar approximate analysis has correctly predicted stable
quasi-2D solitons in dipolar BEC \cite{Tikho}.

Condition (\ref{eq3}) suggests that metastable 3D solitons may exist in
free space when the SOC term is present, while its strength $\lambda $ is
not too large, $N$ and $\eta $ being not too large either. We confirm these
expectations below by means of accurate numerical analysis.

\emph{The Gross-Pitaevskii equation ---} Energy functional (\ref{eq1}) gives
rise to the Gross-Pitaevskii equation (GPE) for the spinor wave function,
\begin{gather}
\left[ i\frac{\partial }{\partial t}+\frac{1}{2}\nabla ^{2}+i\lambda \nabla
\cdot {\boldsymbol{\sigma }}\right.  \notag \\
\left. +g\left(
\begin{array}{cc}
|\psi _{+}|^{2}+\eta |\psi _{-}|^{2} & 0 \\
0 & |\psi _{-}|^{2}+\eta |\psi _{+}|^{2}%
\end{array}%
\right) \right] \Psi =0.  \label{GPE}
\end{gather}%
Assuming axial symmetry of the expected self-trapped states (it is the
highest symmetry admitted by the SOC \cite{Sakaguchi}) and using cylindrical
coordinates $(r,z,\phi )$, the stationary wave function with integer
vorticity $m\geq 0$ and chemical potential $\mu $ is looked for as
\begin{equation}
\left(
\begin{array}{c}
\psi _{+} \\
\psi _{-}%
\end{array}%
\right) =e^{-i\mu t}\left(
\begin{array}{c}
e^{im\phi }\,f_{1}(r,z) \\
e^{i(m+1)\phi }\,f_{2}(r,z)%
\end{array}%
\right) \,.\,  \label{eq5}
\end{equation}%
%
%
%
%
%
%
%
%
Following the terminology introduced for 2D solitons in Ref.~\cite{Sakaguchi}%
, self-trapped states (\ref{eq5}) with $m=0$ are called \textit{semi-vortices%
} (SVs), the states with $m\geq 1$ being their excited states. Similar to
the 2D system \cite{Sakaguchi}, our calculations demonstrate that
the energy of the SV with $m=0$ is always lowest, therefore we focus on $m=0$%
.


Due to the up-down symmetry of underlying Hamiltonian (\ref{eq1}),
degenerate to SV (\ref{eq5}) is its flipped counterpart,
\begin{equation}
\left(
\begin{array}{c}
\psi _{+} \\
\psi _{-}%
\end{array}%
\right) = e^{-i\mu t} \,\left(
\begin{array}{c}
e^{-i\left( m+1\right) \phi }\,f_{2}^{\ast }(r,z) \\
e^{-im\phi }\,f_{1}^{\ast }(r,z)%
\end{array}%
\right) \,,\,  \label{flipped}
\end{equation}%
with $\ast $ standing for the complex conjugate. Although the system is
axially symmetric, stationary states do not necessarily follow this
symmetry. In particular, any superposition of \textit{ans\"{a}tze}~(\ref{eq5}%
) and (\ref{flipped}) breaks the symmetry. Following the nomenclature
introduced in Ref. \cite{Sakaguchi}, we call the state generated by such a
superposition a \textit{mixed mode }(MM). Approximating it by the
superposition with mixing angle $\theta $ \cite{note},%
\begin{equation}
\begin{split}
\psi _{+}& =\left( \cos {\theta }\right) \,f_{1}(r,z)-\left( \sin {\theta }%
\right) \,f_{2}^{\ast }(r,z)\,e^{-i\phi }\,, \\
\psi _{-}& =\left( \sin {\theta }\right) \,f_{1}^{\ast }(r,z)+\left( \cos {%
\theta }\right) \,f_{2}(r,z)\,e^{i\phi }\,,
\end{split}
\label{eq10}
\end{equation}%
straightforward calculation relates its energy to that of the respective SV:
\begin{eqnarray}
E_{\mathrm{MM}} &=&E_{\mathrm{SV}}+(1-\eta )\sin ^{2}{\theta }\cos ^{2}{%
\theta }\,\Delta E\,,  \label{eq12} \\
\Delta E &=&2\pi g\int \!rdr\int \!dz\left(
|f_{1}|^{4}+|f_{2}|^{4}-4|f_{1}|^{2}|f_{2}|^{2}\right) .  \notag
\end{eqnarray}%
Our numerical calculations show that $\Delta E$ is always positive,
hence, like in the 2D case \cite{Sakaguchi}, the SV (MM) has lower energy at
$\eta <1$ ($\eta >1$) . This prediction is confirmed below by the full
numerical analysis.

\begin{figure}[tbh]
\centering\includegraphics[width=1\columnwidth]{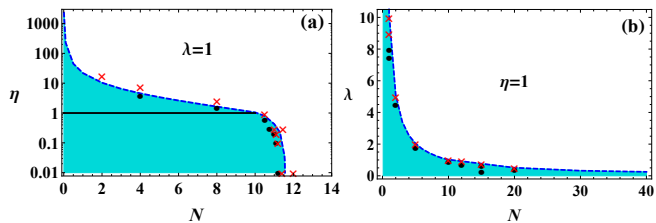}
\caption{(Color online) 3D stable solitons are predicted by the variational
calculation in blue shaded regions of the respective parameter planes. In
panel (a), these are SVs (semi-vortices) at $\protect\eta <1$, and MMs
(mixed modes) at $\protect\eta >1$, with the boundary between them depicted
by the black solid line. In (b), the entire stability area is filled by the
solitons of both types, as the SVs and MMs have equal energies at $\protect%
\eta =1$. The predictions are accurately confirmed by full numerical
simulations, as indicated by red crosses and black dots, which indicate,
respectively, the absence and presence of stable solitons for respective
sets of parameters. }
\label{fig2}
\end{figure}

\begin{figure}[tbh]
\centering
\includegraphics[width=0.95\columnwidth]{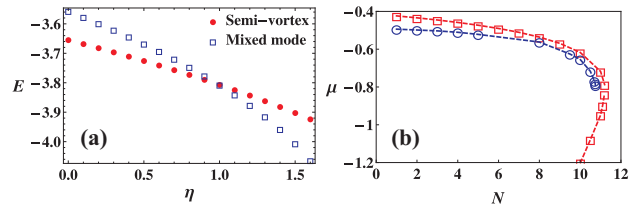}
\caption{(Color online) (a) Energies of the SVs and MMs, as predicted by the
variational approach for $g=\protect\lambda =1$ and $N=8$. The two curves
cross at $\protect\eta =1$, where the SV and MM have equal energies. (b) The numerically (blue circles) and variationally
(red squares) found chemical potential vs. the norm for the SVs at $g=%
\protect\lambda =1$ and $\protect\eta =0.3$. The numerical branch extends 
up to $N=N_{\max }$, in agreement with Eq.~(\ref{NN}).}
\label{fig3}
\end{figure}

\emph{Variational analysis ---} To produce analytical results in a more
accurate form than given by Eq. (\ref{eq2}), we here adopt the following
ansatz for the SV:
\begin{equation}
f_{n}=i^{n-1}\left( A_{n}+iB_{n}z\right) r^{n-1}e^{-\alpha _{n}r^{2}-\beta
_{n}z^{2}}\;\;(n=1,2)\,,  \notag
\end{equation}%
with real parameters $A_{n},B_{n}$, and $\alpha _{n}>0$, $\beta _{n}>0$. The
substitution of this ansatz into expression (\ref{eq1}) for the full energy
and minimizing it with respect to the free parameters produces algebraic
equations which can be readily solved numerically. Stable solitons
correspond to finite values of $\alpha _{n}$ and $\beta _{n}$, while $\alpha
_{n},\beta _{n}\rightarrow 0$ (spreading) and $\alpha _{n},\beta
_{n}\rightarrow \infty $ (collapsing) indicate that no solitons exist.
Results of the calculations are summarized in Fig.~\ref{fig2}, in which the
stable 3D solitons are predicted to exist in the shaded areas. We thus
conclude that the solitons indeed exist, provided that $\lambda $, $N$ and $%
\eta $ are not too large, in agreement with the qualitative prediction of
Eq.~(\ref{eq3}) from the dimensional analysis. In particular, an important conclusion is that, for fixed $%
\lambda $ and $\eta $, the stable solitons always exist in a finite interval
of the norm,
\begin{equation}
0\leq N\leq N_{\max }\left( \lambda ,\eta \right)\, .  \label{NN}
\end{equation}%
Furthermore, as shown in Fig.~\ref{fig3}(a), for $\eta <1$ the energy of the SV
predicted by the variational analysis (VA) is lower than that for the MM,
and vice versa for $\eta >1$, in agreement with the prediction of Eq.~(\ref%
{eq12}).

The red squares in Fig.~\ref{fig3}(b) represent the variational results for the soliton's chemical potential, $\mu $, plotted as 
a function of norm $N$ for $g=\lambda =1$ and $\eta =0.3$. In agreement with the analytical
prediction given by Eq.~(\ref{NN}), there is no threshold (minimum norm)
necessary for the appearance of the solitons, which exist up to a
$N=N_{\max }$. Furthermore, the negative slope of the
dependence, $d\mu /dN<0$, of the upper branch is an indication of the
stability of the soliton families, pursuant to the Vakhitov-Kolokolov (VK)
criterion \cite{Collapse1,Sakaguchi,VK}. The lower branch, which does not
satisfy the VK criterion, represents solitons corresponding to the energy
maximum on the blue dashed curve in Fig.~\ref{fig1}. In the limit of $\mu \to -\infty$, they carry
over into the well-known strongly unstable 3D solitons of the GPE \cite%
{Silberberg}.

\begin{figure}[tbh]
\centering
\includegraphics[width=0.75\columnwidth]{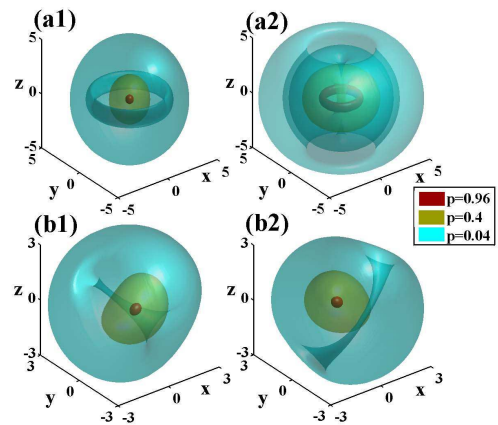}
\caption{(Color online) Density profiles of 3D solitons for $N=8$ and $g=%
\protect\lambda =1$. (a) An SV for $\protect\eta =0.3$, whose fundamental
and vortical components, $|\protect\psi _{+}|$ and $|\protect\psi _{-}|$,
are plotted in (a1) and (a2), respectively. (b) An MM for $\protect\eta =1.5$%
, with (b1), (b2) displaying $|\protect\psi _{+}|$ and $|\protect\psi _{-}|$%
, respectively. In each subplot, different colors represent
constant-magnitude surfaces, $|\protect\psi _{\pm }|=\left(
0.96,0.4,0.04\right) \times |\protect\psi _{\pm }|_{\mathrm{max}}$.}
\label{fig4}
\end{figure}

\emph{Full numerical calculations ---} The prediction for the existence of
the stable 3D solitons in free space, provided by the analytical
approximations, calls for verification by direct simulations of GPE (\ref%
{GPE}). First, we generated stationary states by running the simulations in
imaginary time. Typical examples of the so produced SV and MM density
profiles are displayed in Fig.~\ref{fig4}. Symbols in Fig.~\ref{fig2}, which
indicate the absence and presence of stable solitons, are in good agreement
with the VA.

The blue circles in Fig.~\ref{fig3}(b) represent the numerically obtained chemical potential, which are in good agreement 
with the prediction of the VA. The unstable branch from the VA, however, cannot be produced by the
imaginary-time integration. We have verified the stability of the solitons belonging to the upper branch
in Fig.~\ref{fig3}(b) by real-time simulations with random perturbations added
to the initial conditions, confirming that the VA accurately predicts the SV and MM
stability areas which are displayed in Figs.~\ref{fig2} and \ref{fig3}.

\begin{figure}[tbh]
\centering
\includegraphics[width=0.9\columnwidth]{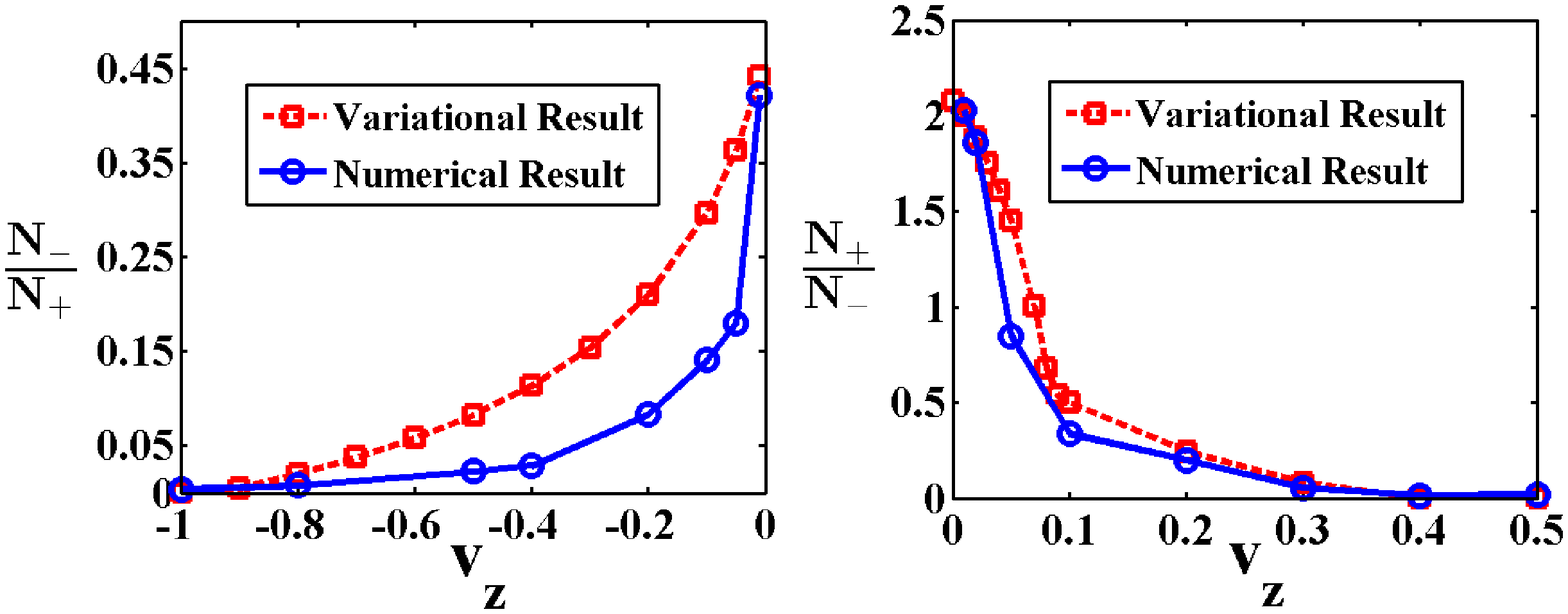}
\caption{(Color online) The ratio of the spin populations as a function of
velocity $v_{z}$ for the moving SV with $N=8$, $g=%
\protect\lambda =1$, $\protect\eta =0.3$, and $N_{\pm }\equiv \protect\int %
d^{3}{r}\,|\protect\psi _{\pm }(\mathbf{r})|^{2}$. The red dashed
lines with squares are variational results, while the blue solid lines with
circles are obtained numerically, using the imaginary-time integration in
the moving reference frame.}
\label{fig6}
\end{figure}

Setting quiescent solitons in motion is another nontrivial issue, as the SOC
terms break the Galilean invariance of the system. To construct solitons
moving along the $z$-axis with velocity $v_{z}$, so that $\psi _{\pm }=\psi
_{\pm }(r,\phi ,z-v_{z}t,t)$, we have rewritten the GPE system (\ref{GPE})
in the respective moving reference frame.
In this form, the velocity term affects the SOC strength along the $z$ axis,
breaking the symmetry between the two components of the spinor. As a result,
positive (negative) $v_{z}$ tends to increase the population of the
spin-down (-up) component. In Fig.~\ref{fig6}, we plot the ratio of the spin
populations as a function of $v_{z}$. Both VA and numerical results are
displayed, showing qualitatively similar results. At $v_{z}<-0.9$ and $%
v_{z}>+0.4$, the moving semi-vortex practically degenerates into a
single-component soliton -- the fundamental or vortical one, respectively --
thus reducing the setting to that for the single GPE with the cubic
self-attraction, where all 3D solitons are strongly unstable. Consequently,
the speed of the stably moving solitons cannot be too large.

\begin{figure}[tbh]
\centering
\includegraphics[width=1.0\columnwidth]{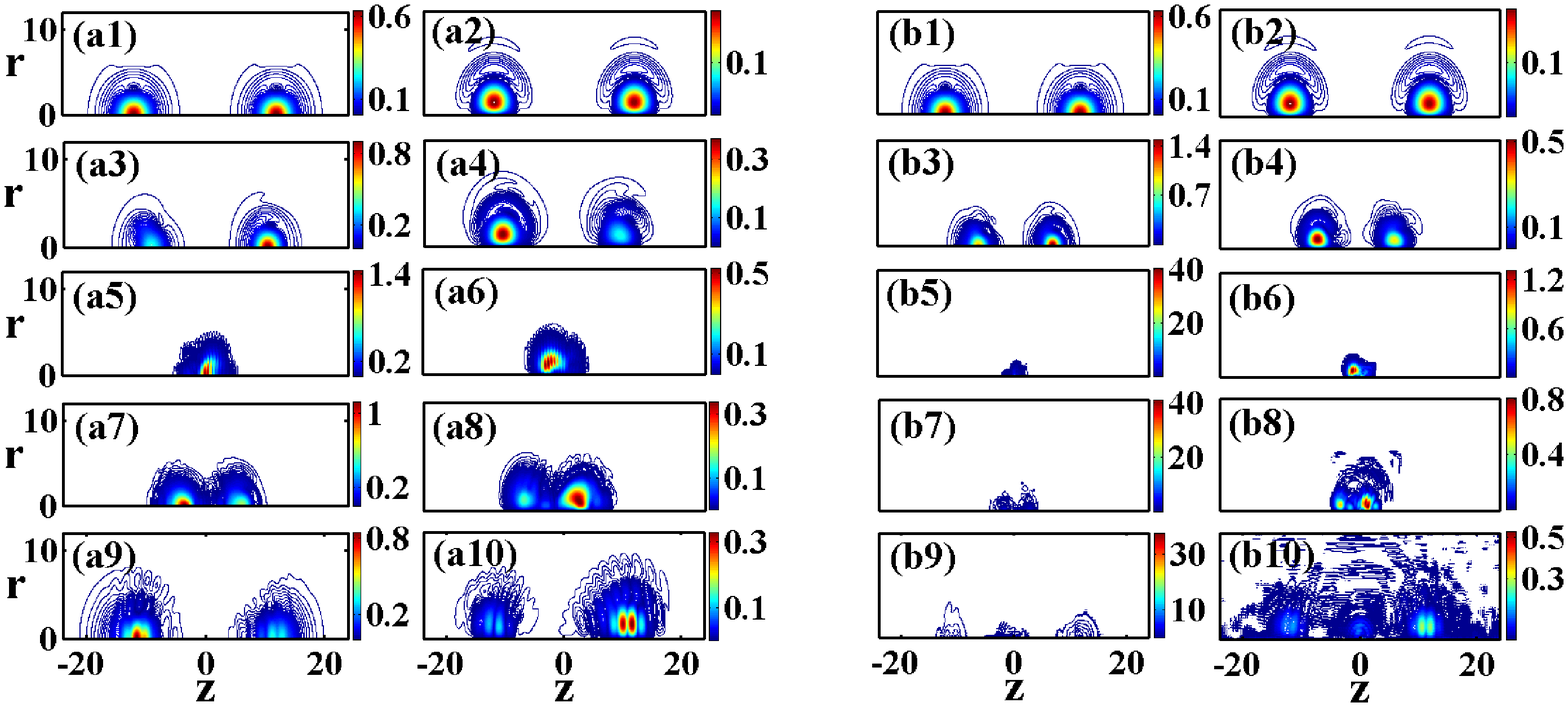}
\caption{(Color online) Collisions of stable 3D SVs in the harmonic trap for
$N=8$, $g=\protect\lambda =1$, and $\protect\eta =0.3$. Panels (a1, a2),
(a3, a4), (a5, a6), (a7, a8) and (a9, a10) display density distributions for
$\Omega =0.5$ at $t=0$, $1.2$, $3.2$, $4$ and $6$, respectively. Panels (b1,
b2), (b3, b4), (b5, b6), (b7, b8) and (b9, b10) display the distributions
for $\Omega =1$ at $t=0$, $1$, $1.4$, $2$ and $3$, respectively. In all
panels, the left and right subplots display, severally, $|\protect\psi _{+}|$
and $|\protect\psi _{-}|$. }
\label{fig7}
\end{figure}

Finally, to consider collisions between moving solitons, we place
two solitons centered at initial positions $(r,z)=(0,\pm z_{0})$, and
include a trapping potential, $\Omega ^{2}(r^{2}+z^{2})/2$. The solitons
then start moving to collide at the trap center, with the trapping frequency
$\Omega $ used to control the collision velocity. Figure~\ref{fig7} depicts
two collision events for the same initial soliton pair. In panel (a), the
slowly moving solitons feature a quasi-elastic collision, while in (b) the
collision leads to destruction of faster solitons. This shows the solitons
are robust against slow collisions.

\emph{Conclusion ---} The combination of the analytical and numerical
methods reveals that stable free-space 3D solitons can be supported in the
binary atomic condensate with attractive interactions and properly
engineered SOC, notwithstanding the presence of the supercritical collapse
in the same setting. This is the first example of metastable solitons in the
3D homogeneous environment with local cubic self-attraction, which exist in
spite of the \emph{nonexistence} of the GS (ground state) in the system. The
SOC plays a crucial role for the stabilization, altering the energy of the
self-trapped states so as to create the local energy minimum. This is the
fundamental difference from the recently discovered stabilization mechanism
in 2D \cite{Sakaguchi}, which readily creates a missing GS below the
critical value of the norm (at $N<N_{\mathrm{cr}}$), where solitons, if any,
cannot be destabilized by the critical collapse, as it does not occur at $%
N<N_{\mathrm{cr}}$, but no solitons could be created at $N\geq N_{\mathrm{cr}%
}$. In 3D, the existence of the metastable solitons is controlled not by the
norm [in an appropriate parameter region, they can be created for any $N$,
although the appropriate region becomes very narrow for very large $N$, as
seen in Fig. \ref{fig2}(b)], but by the energy, as the above analysis
clearly shows.


Although we have adopted the isotropic SOC term in the Hamiltonian, in the
form of $\lambda \mathbf{p}\cdot {\boldsymbol{\sigma }}$, the stabilization
of the 3D solitons does not critically depend on this form,
additional analysis demonstrating that the metastable 3D solitons
exist as well if the SOC strength is different along different axes. It may 
also be interesting to find out if 3D solitons can be stabilized by 
\emph{spatially localized} SOC (for 1D solitons, this 
setting was studied in Ref.~\cite{localized}, but the stability is not an
issue in that case). Influence of the Zeeman splitting, which breaks the
up-down symmetry of the spinor components, on the stability of the solitons
is another relevant problem for further analysis.

On the experimental side, 2D SOC was recently created in an ultracold Fermi
gas \cite{2d}. Realization of 3D SOC may be expected in the
near future, as there is no fundamental obstacle for doing that.

ZWZ acknowledges support from the "Strategic Priority Research
Program (B)" of the CAS, Grant No. XDB01030200 and National Natural Science
Foundation of China (Grant No. 11574294,11174270). HP
acknowledges support from US NSF and the Welch Foundation (Grant No.
C-1669). BAM appreciates a partial support from the Binational (US-Israel)
Science Foundation (grant No. 2010239), and hospitality of the {Department
of Physics and Astronomy at the Rice University}.


\begin{thebibliography}{99}
\bibitem{dark} We are only concerned with bright solitons supported by
attractive nonlinearity. Dark solitons may exist in systems with repulsive
nonlinearity, but their physical realization always requires some
confinement potential and hence can never exist in free space.
\end{thebibliography}
\end{document}